\documentclass[graybox]{svmult}

\usepackage{graphicx}

\usepackage{amsmath}
\usepackage{color}
\usepackage{cite}

\usepackage[latin1]{inputenc}
\usepackage{amsfonts}
\usepackage{tikz} 
\usetikzlibrary{matrix}
\usepackage[english]{babel}
\usepackage[latin1]{inputenc}
\usepackage[T1]{fontenc}
\usepackage{ae}

\usepackage{url}

\allowdisplaybreaks[4]

\newcommand{\bh}{\mathbb{H}}
\newcommand{\bc}{\mathbb{C}}
\newcommand{\bq}{\mathbb{Q}}
\newcommand{\bz}{\mathbb{Z}}
\newcommand{\slz}{\mathrm{SL}(2,\bz)}
\newcommand{\abcd}{\left(\begin{smallmatrix}a&b\\c&d\end{smallmatrix}\right)}

\usepackage{color}
\usepackage{tikz}
\usetikzlibrary{matrix}
\usepackage{cite}           
\usepackage{mathptmx}       
\usepackage{helvet}         
\usepackage{courier}        
\usepackage{type1cm}        
\usepackage{makeidx}         
\usepackage{graphicx}        
\usepackage{multicol}        
\usepackage[bottom]{footmisc}

\newcommand{\cC}{\begin{cal}C\end{cal}}

\newcommand{\cG}{\begin{cal}G\end{cal}}

\newcommand{\cM}{\begin{cal}M\end{cal}}

\newcommand{\cS}{\begin{cal}S\end{cal}}

\let\Im\relax\DeclareMathOperator{\Im}{Im}

\newtheorem{exemp}{Example}

\makeindex             

\begin{document}

\title*{From modular forms to differential equations for Feynman integrals}
\author{Johannes Broedel${}^{a}$, Claude Duhr${}^{b,c}$, Falko Dulat${}^d$, Brenda Penante${}^b$ and Lorenzo Tancredi${}^b$}
\authorrunning{J.~Broedel, C.~Duhr, F.~Dulat, B.~Penante and L.~Tancredi}
\institute{  
${}^a$ Institut f\"{u}r Mathematik und Institut f\"{u}r Physik, Humboldt-Universit\"{a}t zu Berlin,\\
IRIS Adlershof, Zum Grossen Windkanal 6, 12489 Berlin, Germany\\ 
${}^b$ Theoretical Physics Department, CERN, Geneva, Switzerland\\
${}^c$ Center for Cosmology, Particle Physics and Phenomenology (CP3),\\
Universit\'e Catholique de Louvain, 1348 Louvain-La-Neuve, Belgium\\
${}^d$ SLAC National Accelerator Laboratory, Stanford University, Stanford, CA 94309, USA\\
\email{jbroedel@physik.hu-berlin.de }\\
\email{claude.duhr@cern.ch}\\
\email{dulatf@slac.stanford.edu}\\
\email{b.penante@cern.ch}\\
\email{lorenzo.tancredi@cern.ch}}
%

%
\maketitle
\begin{small}CP3-18-41, CERN-TH-2018-152, HU-Mathematik-2018-07, HU-EP-18/19, SLAC-PUB-17293\end{small}\\

\abstract{
In these proceedings we discuss a representation for modular forms that is more
suitable for their application to the calculation of Feynman integrals in the
context of iterated integrals and the differential equation method. In
particular, we show that for every modular form we can find a representation in
terms of powers of complete elliptic integrals of the first kind multiplied by
algebraic functions. We illustrate this result on several examples. In
particular, we show how to explicitly rewrite elliptic multiple zeta values as
iterated integrals over powers of complete elliptic integrals and rational
functions, and we discuss how to use our results in the context of the system
of differential equations satisfied by the sunrise and kite integrals.
}


%
%

\section{Introduction}
\label{sec:1}

Recently, a lot of progress has been made in understanding elliptic multiple polylogarithms (eMPLs)~\cite{BrownLevin},
and in particular their use in the calculation of multiloop Feynman 
integrals~\cite{Broedel:2017kkb,Broedel:2017siw,Broedel:2018iwv}.
As of today, a clear formulation for these functions is available in two different languages.
The first, as iterated integrals over a set of kernels defined on a torus, is preferred
in the mathematics community and finds natural applications in the calculation of one-loop open-string
scattering amplitudes~\cite{Broedel:2014vla,Broedel:2015hia,Broedel:2017jdo}.
The second, as iterated integrals on an elliptic curve defined as the zero-set of a polynomial equation of degree three or four, is more natural in the context of the calculation of 
multiloop Feynman integrals by direct integration 
(for example over their Feynman-Schwinger parameter representation).
In spite of this impressive progress, it remains not obvious how to connect these two languages to that of the differential equations 
method~\cite{Kotikov:1990kg,Remiddi:1997ny,Gehrmann:1999as,Henn:2013pwa}, which constitutes one of the most powerful tools 
for the computation of large numbers of complicated multiloop Feynman integrals.

It is well known that Feynman integrals fulfil systems of linear differential equations 
with rational coefficients in the kinematical invariants and the dimensional regularization parameter $\epsilon$. Once the differential equations are expanded
in $\epsilon$, a straightforward application of
Euler's variation of constants allows one to naturally write their solutions 
as iterated integrals over rational functions and (products of) their homogeneous solutions.
The homogeneous solutions can in turn be inferred by the study of the maximal cut of the corresponding Feynman integrals~\cite{Primo:2016ebd}
and are in general given by non-trivial transcendental functions of the kinematical invariants.
When dealing with Feynman integrals which evaluate to ordinary multiple polylogarithms (MPLs), 
the homogeneous solutions are expected to be
algebraic functions (or at most logarithms). In the ellipitic case, they are instead given by (products of) complete elliptic integrals~\cite{Laporta:2004rb,Remiddi:2016gno,Bonciani:2016qxi,
 vonManteuffel:2017hms,Primo:2017ipr,Adams:2018bsn,Adams:2018kez}.
The iterated integrals arising naturally from this construction 
have been studied in the literature in different special cases~\cite{Ablinger:2017bjx,Remiddi:2017har}, and are particular instances of 
the `iterative non-iterative integrals' considered in refs.~\cite{Ablinger:2017bjx,Ablinger:2017cin}.
A natural question is how and when these new types of iterated integrals 
can be written in terms of the eMPLs
defined in the mathematical literature.
In other words, is it possible to  phrase the solution of the differential equations
for elliptic Feynman integrals directly in terms of eMPLs, and if yes under which conditions?
An obstacle when trying to address this question is that the kernels defining eMPLs do not present themselves in terms of complete elliptic integrals. 
A first possible hint to an answer to this apparent conundrum comes from the observation that elliptic polylogarithms evaluated at some special
points can always be written as iterated integrals of modular forms~\cite{Broedel:2018izr}, and a representation of the equal-mass sunrise 
in terms of this class of iterated integrals also exists~\cite{Adams:2017ejb,Adams:2018yfj,Broedel:2018izr}. 
It is therefore tantalising to speculate that the new class of iterated integrals showing up in Feynman integrals are closely connected to iterated integrals of modular forms and generalisations thereof.

In these proceedings, we start investigating the fascinating problem of how to relate
iterated integrals of modular forms to iterated integrals over rational/algebraic functions and products of
complete elliptic integrals. We mostly focus here on a simpler subproblem, namely on how to express modular forms in
terms of powers of complete elliptic integrals, multiplied by suitable algebraic functions. 
This is a first step towards classifying the new classes of integration kernels that show up in Feynman integral computations, 
and how these new objects are connected to classes of iterated integrals studied in the mathematics literature.
As a main result, we will show that, quite in general, modular forms admit
a representation in terms of
linearly independent products of elliptic integrals and
algebraic functions.
The advantage of this formulation of modular forms (for applications to Feynman integrals) lies in the fact that 
we can describe them in ``purely algebraic terms'', where all quantities are parametrised by variables 
constrained by polynomial equations --  a setting more commonly encountered in physics problems than 
the formulation in terms of modular curves encountered in the mathematics (and string theory) literature.
At the same time, since this formulation is purely algebraic, it lends itself 
more directly to generalisations to cases that cannot immediately be matched to the mathematics of modular forms, e.g., in cases of Feynman integrals depending on more than one kinematic variable.

This contribution to the proceedings is organised as follows: in section \ref{sec:2} we provide a
brief survey of the necessary concepts such as congruence subgroups of $\slz$,
modular forms, Eisenstein and cuspidal subspaces and modular curves.  Section
\ref{sec:3} contains the main part of our contribution: we will show that one
can indeed find suitable one-forms in an algebraic way, which we demonstrate to be in
one-to-one correspondence with a basis of modular forms. 
Finally, we briefly
discuss three applications in section~\ref{sec:application} and present our conclusions in section~\ref{sec:X}.

\vspace{1mm}\noindent



\section{Terms and definitions}
\label{sec:2}

\subsection{The modular group $\slz$ and its congruence subgroups}
In these proceedings we are going to consider functions defined on the
extended upper half-plane $\overline{\bh} = \bh \cup \bq \cup \{i\infty\}$,
where \mbox{$\bh = \{\tau \in \bc\, |\, \Im \tau >
0\}$}.  The modular group $\slz$ acts on the points in $\overline{\bh}$ through
M\"obius transformations of the form
\begin{equation}
  \label{eq:modulartrafo}
  \gamma \cdot \tau = \frac{a\tau + b}{c\tau + d},\quad \gamma = \abcd\in\slz.
\end{equation}
In the following, we will be interested in subgroups of the full modular group.
Of particular interest are the so-called \emph{congruence subgroups of level}
$N$ of $\slz$, 
\begin{align}
\label{eq:subgroups}
\begin{split}
  \Gamma_0(N) &= \{\abcd \in \slz \, :\,c=0\,\bmod\,N\}\,,\\
  \Gamma_1(N) &= \{\abcd \in \slz \, :\,c=0\,\bmod\,N \text{ and }a=d=1\bmod N\}\,,\\
  \Gamma(N)   &= \{\abcd \in \slz \, : \, b=c=0\,\bmod N\,\text{ and }a=d=1\bmod N\}\,.
\end{split}
\end{align}
It is easy to see that $\Gamma\subseteq\slz$ acts separately on $\bh$ and
$\bq\cup\{i\infty\}$.  The action of $\Gamma$ decomposes $\bq\cup\{i\infty\}$
into disjoint orbits.  We refer to the elements of the coset-space
$(\bq\cup\{i\infty\})/\Gamma$ (i.e., the space of all orbits) as \emph{cusps}
of $\Gamma$.  By abuse of language, we usually refer to the elements of the
orbits also as cusps.  We note here that the number of cusps is always finite
for any of the congruence subgroups considered in eq.~\eqref{eq:subgroups}.

\begin{exemp}
One can show that for every rational number $\frac{a}{c} \in \bq$, there is a
matrix $\left(\begin{smallmatrix} a& b\\c& d\end{smallmatrix}\right)\in \slz$
such that $\frac{a}{c} = \lim_{\tau\to i\infty} \frac{a\tau+b}{c\tau+d}$.
Hence, under the action of the group  $\Gamma(1)\simeq\slz$ every rational
number lies in the orbit of the point $i\infty$, and so $\Gamma(1)$ has a
single cusp which we can represent by the point $i\infty\in \overline{\bh}$,
often referred to as the {cusp at infinity}.
 
At higher levels a congruence subgroup usually has more than one cusp. 
For example, the group $\Gamma(2)$ has three cusps, which we may represent by 
$\tau=i\infty$, $\tau=0$ and $\tau=1$. Representatives for the cusps of congruence
subgroups of general level~$N$ can be obtained from {SAGE}~\cite{sage}. 
\end{exemp}


\subsection{Modular curves}
\label{ssec:modcurve}
Since the action of any congruence subgroup $\Gamma$ of $\slz$ allows us to
identify points in the (extended) upper half-plane $\bh$ ($\overline{\bh}$), it
is natural to consider its quotient by $\Gamma$, commonly referred to as a
\emph{modular curve},
\begin{equation}
        X_{\Gamma} \equiv \overline{\bh}/\Gamma \quad \textrm{ and }\quad  Y_{\Gamma} \equiv \bh/\Gamma\,.
\end{equation}
In the cases where $\Gamma$ is any of the congruence subgroups in
eq.~\eqref{eq:subgroups}, the corresponding modular curves are usually denoted
by $X_0(N) \equiv X_{\Gamma_0(N)}$, $X_1(N) \equiv X_{\Gamma_1(N)}$ and $X(N)
\equiv X_{\Gamma(N)}$.

There is a vast mathematical literature on modular curves, and we content
ourselves here to summarise the main results which we will use in the remainder
of these proceedings.  It can be shown that $Y_{\Gamma}$ always defines a Riemann
surface, which can be compactified by adding a finite number of points to
$Y_{\Gamma}$, which are precisely the cusps of $\Gamma$. In other words, while
$Y_{\Gamma}$ is in general not compact, $X_{\Gamma}$ always defines a
\emph{compact} Riemann surface. Hence, we can apply very general results from
the theory of compact Riemann surfaces to the study of modular curves, as we
review now.

First, every (compact) Riemann surface can be explicitly realised as the
zero-set of a polynomial $\Phi(x,y)$ in two variables.\footnote{More
rigorously, one should consider the zero set a homogeneous polynomial
$\Phi(x,y,z)$ in $\mathbb{CP}^2$. For simplicity, we will always work here in
the affine chart $z=1$ of $\mathbb{CP}^2$.} In other words, we have (at least
in principle) two ways to describe the modular curve $X_\Gamma$: either as the
quotient  of the extended upper half plane, or as the projective curve $\cC$ in
$\mathbb{CP}^2$ defined by the polynomial equation $\Phi(x,y)=0$. Hence, there
must be a map from $\overline{\bh}/\Gamma$ to $\cC$ which assigns to
$\tau\in\overline{\bh}/\Gamma$ a point $(x(\tau),y(\tau))\in\cC$ such that
$\Phi(x(\tau),y(\tau))=0$. Since two points in $\overline{\bh}/\Gamma$ are
identified if they are related by a M\"obius transformation for $\Gamma$, the
functions $x(\tau)$ and $y(\tau)$ must be invariant under modular
transformations for $\Gamma$, e.g.,
\begin{equation}\label{eq:modular_function}
x\left(\frac{a\tau+b}{c\tau+d}\right) = x(\tau)\,,\qquad \forall \left(\begin{smallmatrix} a& b\\c& d\end{smallmatrix}\right)\in \Gamma\,,
\end{equation}
and similarly for $y(\tau)$. A meromorphic function satisfying
eq.~\eqref{eq:modular_function} is called a \emph{modular function} for
$\Gamma$. Equivalently, the modular functions for $\Gamma$ are precisely the
meromorphic functions on $X_\Gamma$. Note that since $X_{\Gamma}$ is compact,
there are no non-constant holomorphic functions on $X_{\Gamma}$ (because they
would necessarily violate Liouville's theorem). Modular functions can easily be
described in terms of the algebraic curve $\cC$: they are precisely the
rational functions in $(x,y)$ subject to the constraint $\Phi(x,y)=0$.
Equivalently, the field of modular functions for $X_{\Gamma}$ is the field
$\mathbb{C}(x(\tau),y(\tau))$. In particular, we see that the field of
meromorphic functions of a modular curve (or of any compact Riemann surface)
has always (at most) two generators $x$ and $y$.

\begin{exemp}
It can be shown that the modular curve $X_0(2)$ is isomorphic to the algebraic variety $\cC$ described by the zero-set of the polynomial
\begin{equation}\begin{split}\label{eq:X02_example}
\Phi_2(x,y) &\,= x^3 +y^3 -162000 (x^2+y^2) + 1488 x y (x+y) -x^2y^2 + 8748000000(x+y) \\
&\,+ 40773375 xy -157464000000000\,.
\end{split}\end{equation}
In general, the coefficients of the polynomials describing modular curves are very large numbers, already for small values of the level $N$. The map from the quotient space $\overline{\bh}/\Gamma_0(2)$ to the curve $\cC$ is given by\footnote{The notation $j'(\tau)\equiv j(2\tau)$ is standard in this context in the mathematics literature, though we emphasise that $j'(\tau)$ \emph{does not} correpsond to the derivative of $j(\tau)$.} 
\begin{equation}
\tau \mapsto (x,y) = (j(\tau),j'(\tau)) \equiv (j(\tau),j(2\tau))\,,
\end{equation}
where $j: \bh\to \mathbb{C}$ denotes Klein's $j$-invariant. The field of meromorphic functions of $X_0(2)$ is the field of rational functions in two variables $(x,y)$ subject to the constraint $\Phi_2(x,y)=0$, or equivalently the field $\bc(j(\tau),j'(\tau))$ of rational functions in $(j(\tau),j'(\tau))$.

In general, the polynomials $\Phi_N(x,y)$ describing the \emph{classical modular curves} $X_0(N)$ can be constructed explicitly, cf. e.g. ref.~\cite{X0N_prime,X0N_composite}, and they are available in computer-readable format up to level 300~\cite{X0N_web}. The zeroes of $\Phi_N(x,y)$ are parametrised by $(j(\tau),j'(\tau)) \equiv (j(\tau),j(N\tau))$, the field of meromorphic functions is $\bc(j(\tau),j'(\tau))$.
\end{exemp}

In some cases it is possible to find purely rational solutions to the polynomial equation $\Phi(x,y)=0$, i.e., one can find rational functions $(X(t), Y(t))$ such that $\Phi(X(t),Y(t))=0$ for all values of $t \in \widehat{\bc} \equiv \bc\cup\{\infty\}$. In such a scenario we have constructed a map from the Riemann sphere $\widehat{\bc}$ to the curve $\cC$, and so we can identify the curve $\cC$, and thus the corresponding modular curve $X_{\Gamma}$, with the Riemann sphere. By a very similar argument one can conclude that there must be a modular function $t(\tau)$ for $\Gamma$ which allows us to identify the quotient $\overline{\bh}/\Gamma$ with the Riemann sphere. Such a modular function is called a \emph{Hauptmodul} for $\Gamma$. It is easy to see that in this case the field of meromorphic functions reduces to the field $\bc(t(\tau))$ of rational functions in the Hauptmodul, in agreement with the fact that the meromorphic functions on the Riemann sphere are precisely the rational functions.

\begin{exemp}
It is easy to check that eq.~\eqref{eq:X02_example} admits a purely rational solution of the form~\cite{Maier}
\begin{equation}
(x,y) = (X(t),Y(t)) = \left(\frac{(t+16)^3}{t}, \frac{(t+256)^3}{t^2}\right)\,.
\end{equation}
We have thus constructed a map from the Riemann sphere to the modular curve $X_0(2)$, and so $X_0(2)$ is a curve of genus zero. 
A Hauptmodul for $X_0(2)$ can be chosen to be~\cite{Maier}
\begin{equation}\label{eq:t2_def}
t_2(\tau) = 2^{12}\,\left(\frac{\eta(2\tau)}{\eta(\tau)}\right)^{24}\,,
\end{equation}
where $\eta$ denotes Dedekind's $\eta$-function. 
\end{exemp}

It is possible to compute the genus of a modular curve. In particular, it is
possible to decide for which values of the level $N$ the modular curves
associated to the congruence subgroups in eq.~\eqref{eq:subgroups} have genus
zero. Here is a list of results:
%
\begin{itemize}
\item $X_0(N)$ has genus 0 iff $N \in \{1,\dots,10,12,13,16,18,25\}$.
\item $X_1(N)$ has genus 0 iff $N \in \{1,\dots,10,12\}$.
\item $X(N)$ has genus 0 iff $N \in \{1,2,3,4,5\}$.
\end{itemize}
Hauptmodule for these modular curves have been studied in the mathematics
literature. In particular, the complete list of Hauptmodule for the modular
curves $X_0(N)$ of genus zero can be found in ref.~\cite{Maier} in terms of
$\eta$-quotients.  Other cases are also known in the literature, but they may
involve Hauptmodule that require generalisations of Dedekind's $\eta$-function,
see e.g. ref.~\cite{YifanYang}.

\begin{exemp}\label{exemp:X(2)}
The modular curves $X(1)$ and $X(2)$ have genus zero, and the respective Hauptmodule are Klein's $j$-invariant $j(\tau)$ and the {modular $\lambda$-function},
\begin{equation}\label{eq:modular_lambda}
\lambda(\tau) = \theta_2^4(0,\tau)/\theta_3^4(0,\tau) = 2^4\,\left(\frac{\eta(\tau/2)\,\eta(2\tau)^2}{\eta(\tau)^3}\right)^8\,,
\end{equation}
where $\theta_n(0,\tau)$ are Jacobi's $\theta$-functions.
\end{exemp}


\subsection{Modular forms}
\label{ssec:wmfmf}
One of the deficiencies when working with modular curves is the absence of holomorphic modular functions on $X_{\Gamma}$. 
We can, however, introduce a notion of holomorphic functions by
 relaxing the condition on how the functions should transform under $\Gamma$.
For every non-negative
integer $k$, we can define an action of $\Gamma$ on functions on
$\overline{\bh}$ by
\begin{equation}
  (f|_{k}\gamma)(\tau) \equiv (c\tau+d)^{-k}f(\gamma\cdot\tau)\,,\quad\gamma=\abcd\in\Gamma\,.
\end{equation}
A meromorphic function $\overline{\bh}\to\bc$ is called \emph{weakly modular
of weight $k$} for $\Gamma$ if it is invariant under this action,
\begin{equation}
  \label{eqn:modtrans}
  (f|_{n}\gamma)(\tau) = f(\tau).
\end{equation}
Note that weakly modular functions of weight zero are precisely the modular functions for $\Gamma$.

A \emph{modular form} of weight $k$ for $\Gamma$ is, loosely speaking, a weakly modular function of weight $k$
 that is holomorphic on $\overline{\bh}$. In particular it is
holomorphic at all the cusps of $\Gamma$.  We denote the $\bq$-vector space of
modular forms of weight $k$ for $\Gamma$ by $\mathcal{M}_{k}(\Gamma)$. It can
be shown that this space is always finite-dimensional. We summarise here some properties of spaces of modular forms that are easy to prove and that will be useful later on.
\begin{enumerate}
\item The space of all modular
forms is a graded algebra,
\begin{equation}
  \mathcal{M}_{\bullet}(\Gamma)=\bigoplus_{k=0}^{\infty}\mathcal{M}_{k}(\Gamma), \quad \textrm{with}\quad\mathcal{M}_{k}(\Gamma)\cdot\mathcal{M}_{\ell}(\Gamma)\subseteq\mathcal{M}_{k+\ell}(\Gamma).
\end{equation}
\item If $\Gamma'\subseteq\Gamma$, then $\cM_k(\Gamma)\subseteq\cM_k(\Gamma')$. 
\item If $\left(\begin{smallmatrix}-1&0\\0&-1\end{smallmatrix}\right)\in\Gamma$, then there are no modular forms of odd weight for $\Gamma$.
\end{enumerate}

A modular form that vanishes at all cusps of $\Gamma$ is called a \emph{cusp
form}. The space of all cusp forms of weight $k$ for $\Gamma$ is denoted by
$\cS_k(\Gamma)$. The space of all cusp forms
$\cS_{\bullet}(\Gamma)=\bigoplus_{k=0}^{\infty}\cS_k(\Gamma)$ is obviously
a graded subalgebra of $\mathcal{M}_{\bullet}(\Gamma)$ and an ideal in
$\mathcal{M}_{\bullet}(\Gamma)$. The quotient space is the \emph{Eisenstein
subspace}:
\begin{equation}
  \mathcal{E}_{\bullet}(\Gamma)\simeq\mathcal{M}_{\bullet}(\Gamma)/\cS_{\bullet}(\Gamma).
\end{equation}
Note that at each weight the dimension of the Eisenstein subspace for $\Gamma$ is equal\footnote{There are exceptions for small values of the weight and the level.} to the number of cusps of $\Gamma$.

\begin{exemp} 
Let us analyse modular forms for $\Gamma(1)\simeq SL(2,\bz)$. There are no modular forms for $\Gamma(1)$ of odd weight. Since $\Gamma(1)$ has only one cusp, there is one Eisenstein series for every even weight, the {Eisenstein series $G_{2m}$},
\begin{equation}\label{eq:G_def}
  G_{2m}(\tau) = \sum_{(\alpha,\beta)\in\bz^2\setminus\{(0,0)\}}\frac{1}{(\alpha+\beta\tau)^{2m}}.
\end{equation}
It is easy to check that $G_{2m}(\tau)$ transforms as a modular form of weight $2m$, except when $m=1$, which will be discussed below. The first cusp form for $\Gamma(1)$ appears at weight 12, known as the {modular discriminant},
\begin{equation}\label{eq:modular_disc}
\Delta(\tau) = 2^{12}\,\eta(\tau)^{24} = 10\,800\left(20\,G_4(\tau)^3 - 49\,G_6(\tau)^2\right)\,.
\end{equation}
\end{exemp}

In the same way as the Eisenstein subspace for $\Gamma(1)$ is generated by the
Eisenstein series $G_{2m}(\tau)$, there exist analogues for the Eisenstein
subspaces for congruence subgroups. 

$G_2(\tau)$ is an example of a quasi modular form. A \emph{quasi modular form
of weight $n$ and depth $p$ for $\Gamma$} is a holomorphic function
$f:\overline{\bh}\to\bc$ that transforms
as,
\begin{equation}\label{eq:quasi_modular}
  (f|_n\gamma)(\tau) = f(\tau) + \sum_{r=1}^{p}f_r(\tau)\left(\frac{c}{c\tau+d}\right)^r\,,\quad\gamma=\abcd\in\Gamma\,,
\end{equation}
where $f_1,\dots,f_p$ are holomorphic functions.  In the case of the Eisenstein
series $G_2(\tau)$ we have,
\begin{equation}\label{eq:G2_transform} 
  G_2\left(\frac{a\tau+b}{c\tau+d}\right)=(c\tau+d)^2\Big(G_2(\tau)-\frac{1}{4\pi i}\frac{c}{c\tau+d}\Big).
\end{equation}
Comparing eq.~\eqref{eq:G2_transform} to eq.~\eqref{eq:quasi_modular}, we see that $G_2(\tau)$ is a quasi-modular form of weight two and depth one.

It is easy to check that any congruence subgroup $\Gamma$ of level $N$ contains the element
\begin{equation}
  T^N = \left(\begin{smallmatrix}1&N\\0&1\end{smallmatrix}\right),
\end{equation}
which generates the M\"obius transformation $\tau\to\tau+N$. Consequently,
modular forms of level $N$ are periodic functions
with period $N$ and thus admit Fourier expansions of the form
\begin{equation}
  f(\tau) = \sum_{m=0}^{\infty}a_me^{2\pi im\tau/N}=\sum_{m=0}^{\infty}a_mq_N^m,
\end{equation}
with $q\equiv\exp(2\pi i \tau)$ and $q_N=q^{1/N}$, which are called
\emph{$q$-expansions}.

\begin{exemp}
The Eisenstein series for $\Gamma(1)$ admit the q-expansion
\begin{equation}
  \label{eqn:eis}
G_{2m}(\tau) = 2\zeta_{2m} +\frac{2\,(2\pi i)^{2m}}{(2m-1)!}\sum_{n=1}^\infty\sigma_{2m-1}(n)\,q^n\,,
\end{equation}
where $\sigma_p(n) = \sum_{d|n}d^p$ is the divisor sum function.
\end{exemp}

In the previous section we have argued that modular curves admit a purely algebraic description in terms of zeroes of polynomials in two variables. 
For practical applications in physics such an algebraic description is often desirable, because concrete applications often present themselves in terms of polynomial equations. Such an algebraic description also exists for (quasi-)modular forms. In particular, it was shown by Zagier that every modular form of positive weight $k$ satisfies a linear differential equation of order $k+1$ with algebraic coefficients~\cite{zagiermodular}. More precisely, consider a modular form $f(\tau)$ of weight $k$ for $\Gamma$. We can pick a modular function $t(\tau)$ for $\Gamma$ and locally invert it to express $\tau$ as a function of $t$. Then the function $F(t)\equiv f(\tau(t))$ satisfies a linear differential equation in $t$ of degree $k+1$ with coefficients that are algebraic functions in $t$. In the case where $\Gamma$ has genus zero\footnote{We define the genus of a congruence subgroup $\Gamma$ to be the genus of the modular curve $X_{\Gamma}$.} we can choose $t(\tau)$ to be a Hauptmodul, in which case the coefficients of the differential equation are rational functions. We emphasise that the function $F(t)$ is only defined \emph{locally}, and in general it has branch cuts. 

One of the goals of these proceedings is to make this algebraic description of modular forms concrete and to present a way how it can be obtained in some specific cases. For simplicity we only focus on the genus zero case, because so far modular forms corresponding to congruence subgroups of higher genus have not appeared in Feynman integral computations. We emphasise, however, that this restriction is not essential and it is straightforward to extend our results to congruence subgroups of higher genus.

\vspace{1mm}\noindent



\section{An algebraic representation of modular forms}
\label{sec:3}

\subsection{General considerations}

In this section, we will make the considerations at the end of the previous section concrete, and we are going to construct a basis of modular forms of given
weight for different congruence subgroups of $\slz$ in terms of objects that admit a purely algebraic
description. More precisely, consider a modular form $f$ of weight $k$ for $\Gamma$, where $\Gamma$ can be any of the congruence subgroups in eq.~\eqref{eq:subgroups}. Then, at least \emph{locally}, we can find a modular function $x(\tau)$ for $\Gamma$ and an \emph{algebraic} function $A$ such that
\begin{equation}\label{eq:general}
f(\tau ) = \textrm{K}(\lambda(\tau))^k \, A(x(\tau))\,,
\end{equation}
where $\lambda$ denotes the modular $\lambda$ function of eq.~\eqref{eq:modular_lambda} and $\textrm{K}$ is the
 complete
elliptic integral
of the first
kind,
\begin{equation}
  \begin{split}
  \textrm{K}(\lambda)&=\int_0^1\frac{1}{\sqrt{(1-t^2)(1-\lambda\,t^2)}}\,dt\,.
  \end{split}
\end{equation}
Note that locally we can write $\lambda$ as an algebraic function of $x$, so that the argument of the complete elliptic integral
can be written as an algebraic function of $x$. Since $\textrm{K}$ satisfies a linear differential equation of order two, it is then easy to see that the right-hand side of eq.~\eqref{eq:general} satisfies a linear differential equation of order ${k+1}$ in $x$ with algebraic coefficients. The existence of the local representation in eq.~\eqref{eq:general} can be inferred from the following very simple reasoning. First, since $\Gamma(N)\subseteq \Gamma_1(N)\subseteq \Gamma_0(N)$ it is sufficient to discuss the case of the group $\Gamma(N)$. Next, let $M=\textrm{lcm}(4,N)$ be the least common multiple of $4$ and $N$. Since $\Gamma(M)\subseteq \Gamma(N)$, $f$ is a modular form of weight $k$ for $\Gamma(M)$. One can check that $\textrm{K}(\lambda(\tau))$ is a modular form of weight one for $\Gamma(4)$, and therefore also for $\Gamma(M)$. The ratio $f(\tau)/\textrm{K}(\lambda(\tau))^k$ is then a modular form of weight zero for $\Gamma(M)$, and thus a modular function, i.e., an element of the function field $\bc(x(\tau),y(\tau))$ of $\Gamma(M)$. Hence we have
$f(\tau)/\textrm{K}(\lambda(\tau))^k = R(x(\tau),y(\tau))$.
 $y$~is an algebraic function of $x$ (because they are related by the polynomial equation $\Phi(x,y)=0$ that defines $X(M)$), and so we can choose $A(x(\tau)) = R(x(\tau),y(\tau))$ in eq.~\eqref{eq:general}.

While the previous argument shows that a representation of the form~\eqref{eq:general} exists for any modular form of level $N$, finding this representation in explicit cases can be rather hard.
Our goal is to show that often one can find this representation using analytic constraints, which allow us to infer the precise form of the algebraic coefficient $A$. 
We focus here exclusively on congruence subgroups of genus zero, but we expect that similar arguments apply to higher genera.
In the next paragraphs, we are going to describe the general strategy. In subsequent sections
we will illustrate the procedure on concrete examples, namely the congruence subgroups
$\Gamma(2)$ and $\Gamma_0(N)$ for $N\in\{2,4,6\}$, as well as the group $\Gamma_1(6)$ which is relevant for the sunrise graph~\cite{Bloch:2013tra,Adams:2017ejb}. In particular, we will construct an explicit basis of modular forms for these groups for arbitrary weights. 


Assume that we are given a modular form $B(\tau)$ of weight $p$ for $\Gamma$, which we call \emph{seed modular form}
in the following. In the argument at the beginning of this section the seed modular form is $\textrm{K}(\lambda(\tau))$, assuming that $\Gamma$ contains $\Gamma(4)$ as a subgroup. It is however useful to formulate the argument in general without explicit reference to $\textrm{K}(\lambda(\tau))$.
Next, consider a modular form $f(\tau)$ of weight $k$ for $\Gamma$ with $p|k$. Then by an argument very similar to the one presented at the beginning of this section we conclude that there is a modular function $x(\tau)$ for $\Gamma$ and an algebraic function $A(x)$ such that 
\begin{equation}
  A(x(\tau)) = \frac{f(\tau)}{B(\tau)^{k/p}} \,.
\end{equation}
If $\Gamma$ has genus zero and $x$ is a Hauptmodul for $\Gamma$, then the function $A$ is a rational function of $x$.
From now on we assume for simplicity that we work within this setting.
%

Up to now the argument was similar to the one leading to the form~\eqref{eq:general},
and we have not constrained the form of the rational function $A$. We now discuss how this can be achieved.
Being a modular form, $f(\tau)$ needs to be holomorphic
everywhere.  Correspondingly, the rational function $A(x(\tau))$ can have
poles at most for $B(\tau)=0$. In applications, the location of the poles is usually known (see the next sections).
Let us denote them by $\tau_i$, and we set $x_i = x(\tau_i)$ (with $x_i\neq \infty$). We must have
\begin{equation}
A(x) = \frac{P(x)}{\prod_{i}(x-x_i)^{n_i}}\,,
\end{equation}
where $P(x)$ is a polynomial. The degree of $P$ is bounded by analysing the behaviour of the seed modular form at points where $x(\tau)=\infty$, where both $f$ and $B$ must be holomorphic.
Finally, the modular form $f(\tau)$ can be written as
\begin{equation}\label{eq:B_general}
f(\tau) = \frac{B(\tau)^{k/p}}{\prod_{i}(x(\tau)-x_i)^{n_i}}\,\left[d_0+d_1\,x(\tau) + \ldots +d_m\,x(\tau)^m\right]\,,
\end{equation}
where the $d_i$ are free coefficients. In the next sections we illustrate this construction explicitly on the examples of the congruence subgroups $\Gamma(2)$, $\Gamma_0(N)$, $N\in\{2,4,6\}$ and $\Gamma_1(6)$. However, before we do so, let us make a few comments about eq.~\eqref{eq:B_general}.
First, we see that we can immediately recast eq.~\eqref{eq:B_general} in the form~\eqref{eq:general} if we know how to express the seed modular form $B$ in terms of the complete elliptic integral of the first kind. While we do not know any generic way of doing this a priori, in practical applications the seed modular form will usually be given by a Picard-Fuchs equation whose solutions can be written in terms of elliptic integrals. Second, we see that eq.~\eqref{eq:B_general} depends on $m+1$ free coefficients, and so $\textrm{dim}\,\cM_k(\Gamma)=m+1$. Finally, let us discuss how cusp forms arise in this framework. Let us assume that $\Gamma$ has $n_C$ cusps, which we denote by $\tau_r$, $1\le r\le n_C$. For simplicity we assume that $c_r=x(\tau_r)\neq \infty$, though the conclusions will not depend on this assumption. Then $f$ is a cusp form if $f(\tau_r)=0$ for all $1\le r\le n_C$. It can easily be checked that, by construction, the ratio multiplying the polynomial in eq.~\eqref{eq:B_general} can never vanish. Hence, all the zeroes of $f$ are encoded into the zeroes of the polynomial part in eq.~\eqref{eq:B_general}. Therefore $f$ is a cusp form if and only if it can locally be written in the form
\begin{equation}\label{eq:cusp_general}
f(\tau) = \frac{B(\tau)^{k/p}}{\prod_{i}(x(\tau)-x_i)^{n_i}}\,\left[\prod_{\substack{r=1\\ c_r\neq\infty}}^{n_C}(x(\tau)-c_r)\right]\,\left[\sum_{j=1}^{m-n_c-\delta_{\infty}}d_{j}\,x(\tau)^{j}\right]\,,
\end{equation}
with
\begin{equation}
\delta_{\infty} = \left\{\begin{array}{ll}
1\,,&\textrm{ if $c_r=\infty$ for some $r$\,,}\\
0\,,&\textrm{ otherwise}\,.
\end{array}\right.
\end{equation}

%

\subsection{A basis for modular forms for $\Gamma(2)$}
\label{eq:sec_Gamma(2)}

In this section we derive an algebraic representation for all modular forms of
weight $2k$ for the group $\Gamma(2)$, and we present an explicit basis for such modular forms for arbitrary weights. 
As already mentioned in
Example~\ref{exemp:X(2)}, the modular curve $X(2)$ has genus zero and the associated
Hauptmodul is the modular $\lambda$-function. Since $\left(\begin{smallmatrix}-1&0\\0&-1\end{smallmatrix}\right)\in\Gamma(2)$, there are no modular forms of odd weight. 
The group $\Gamma(2)$ has three cusps, which are represented  by
$\tau=i\infty,\,\tau=1$ and $\tau=0$. Under the modular $\lambda$ function the cusps are mapped to
\begin{equation}
  \label{eqn:G2cups}
  \lambda(i \infty)=0\,,\qquad \lambda(0)=1 \,,\qquad \lambda(1)=\infty \,.
\end{equation}

Next, we need to identify our seed modular form. One can easily check that $B(\tau)\equiv\textrm{K}(\lambda(\tau))^2$ is a modular form of weight two for $\Gamma(2)$. If $f$ denotes a modular form of weight $2k$ for $\Gamma(2)$, then we can form the ratio
\begin{equation}\label{eq:R_Gamma(2)}
  R(\lambda(\tau))\equiv \frac{f(\tau)}{B(\tau)^k} = \frac{f(\tau)}{\textrm{K}(\lambda(\tau))^{2k}}\,,
\end{equation}
where $R$ is a rational function in the Hauptmodul $\lambda$.

In order to proceed, we need to determine the pole structure of $R$, or
equivalently the zeroes of the seed modular form $B$, i.e., of the complete
elliptic integral of the first kind. The elliptic integral
$\textrm{K}(\ell)$ has no zeroes in the complex plane.  Furthermore, it is not
difficult to show that $\textrm{K}(\ell)$ behaves like $1/\sqrt{\ell}$ for
$\ell\to\infty$. So the function $B(\tau)$ becomes zero only at
$\lambda(\tau)=\infty$, which corresponds to $\tau=1\!\!\!\mod \Gamma(2)$.  We thus
conclude that $R(\lambda(\tau))$ cannot have poles at finite values of $\lambda(\tau)$, and so it
must be a polynomial.  The degree of the polynomial is bounded by the
requirement that the ratio in eq.~\eqref{eq:R_Gamma(2)} has no pole at
$\tau=1$. Starting from a polynomial ansatz
\begin{equation}
  R(\lambda(\tau))=\sum\limits_{n=0}^m a_n\lambda(\tau)^n
\end{equation}
we find
\begin{equation}
  f(\tau)=\textrm{K}(\lambda(\tau))^{2k}\sum\limits_{n=0}^m a_n\lambda(\tau)^n\,\,\stackrel{\tau\to1}{\sim}\,\,
  \bigg(\frac{1}{\sqrt{\lambda(\tau)}}\bigg)^{2k} a_m\lambda(\tau)^m=
  a_m\lambda(\tau)^{m-k}\,.
  \end{equation}
We see that $f(\tau)$ is holomorphic at $\tau=1$ if and only if the degree of $R$ is at
most $k$. Thus, we can write the most general ansatz for the modular form of
weight $2k$ for $\Gamma(2)$:
\begin{equation}
  \label{eqn:fGam2}
  f(\tau)=\textrm{K}(\lambda(\tau))^{2k}\sum_{n=0}^k c_n\lambda(\tau)^n \,.
\end{equation}
In turn, this allows to infer the dimension of the space of modular forms of
weight $2k$:
\begin{equation}
   \dim\, \cM_{2k}(\Gamma(2))=k+1,\quad k>1\,,
\end{equation}
and we see that the modular forms
\begin{equation}\label{eq:basis_Gamma(2)}
\textrm{K}(\lambda(\tau))^{2k}\,\lambda(\tau)^n\,,\quad 0\le n\le k+1\,,
\end{equation}
form a basis for $\cM_{2k}(\Gamma(2))$.

Finally, let us comment on the space of cusp forms of weight $2k$ for $\Gamma(2)$. Using eq.~\eqref{eq:cusp_general}, we conclude that the most general element of $\cS_{2k}(\Gamma(2))$ has the form
\begin{equation}\label{eq:cusp_Gamma(2)}
\textrm{K}(\lambda(\tau))^{2k}\,\lambda(\tau)\,(1-\lambda(\tau))\,\sum_{n=0}^{k-3} a_n\,\lambda(\tau)^{n}\,.
\end{equation}
  We see that there are $k-2$ cups forms for $\Gamma(2)$ of weight $2k>2$. This number agrees with the data for the dimensions of Eisenstein and cuspidal subspaces delivered by SAGE \cite{sage}. Moreover, we can easily read off a basis of cusp forms for arbitrary weights.

\begin{exemp}
  \label{ex:seven}
Every Eisenstein series for $\Gamma(1)$ (see eq.~\eqref{eq:G_def}) is a modular form for $\Gamma(2)$, and so we can write them locally in the form
\begin{equation}\normalfont\label{eq:Eis_to_G}
G_{2k}(\tau) = \textrm{K}(\lambda(\tau))^{2k}\,\cG_{2k}(\lambda(\tau))\,,\quad k>1\,,
\end{equation}
where $\cG_{2k}(\ell)$ is a polynomial of degree $k$. For example, for low weights we find
\begin{equation}\begin{split}
\cG_4(\ell) &\,= \frac{16}{45}\,(\ell^2-\ell+1)\,,\\
\cG_6(\ell) &\,= \frac{64}{945}\,(\ell-2)(\ell+1)(2\ell-1)\,,\\
\cG_8(\ell) &\,= \frac{256}{4725}\,(\ell^2-\ell+1)^2\,\,.
\end{split}\end{equation}
In this basis the modular discriminant of eq.~\eqref{eq:modular_disc} takes the form
\begin{equation}\normalfont
\Delta(\tau) = 65\,536\,\textrm{K}(\lambda(\tau))^{12}\,\lambda(\tau)^{2}\,(1-\lambda(\tau))^{2}\,,
\end{equation}
in agreement with eq.~\eqref{eq:cusp_Gamma(2)}. Finally, the Eisenstein series of weight two is not modular, so it cannot be expressed in terms of the basis in eq.~\eqref{eq:basis_Gamma(2)}. We note however that one can write
\begin{equation}\normalfont\label{eq:G2_def}
G_2(\tau) = 4\,\textrm{K}(\lambda(\tau))\,\textrm{E}(\lambda(\tau))+\frac{4}{3}\,(\lambda(\tau)-2)\,\textrm{K}(\lambda(\tau))^2\,,
\end{equation}
where $\normalfont \textrm{E}$ denotes the complete elliptic integral of the second kind
\begin{equation}\normalfont
\textrm{E}(\lambda) = \int_0^1dt\,\sqrt{\frac{1-\lambda\,t^2}{1-t^2}}\,.
\end{equation}
\end{exemp}

\subsection{A basis for modular forms for $\Gamma_0(2)$}

In this section we perform the same analysis for the congruence subgroup $\Gamma_0(2)$. 
The analysis will be very similar to
the previous case, so we will not present all the steps in detail. However,
there are a couple of differences which we want to highlight.

We start by reviewing some general facts about $\Gamma_0(2)$. First, there are no modular forms of odd weight.
Second, $\Gamma_0(2)$ has genus zero
(cf.~Section~\ref{ssec:modcurve}), and a Hauptmodul for $\Gamma_0(2)$ is the function $t_2$
defined in eq.~\eqref{eq:t2_def}. 
Since $\Gamma(2)\subseteq\Gamma_0(2)$, the Hauptmodul $t_2$ is a modular function for
$\Gamma(2)$, and so it can be written as a rational function
of $\lambda$, the Hauptmodul for $\Gamma(2)$. Indeed, one finds
\begin{equation}\label{eq:t2_to_lambda}
  t_2(\tau)=16\frac{\lambda(\tau)^2}{1-\lambda(\tau)}\,.
\end{equation}
Inverting the previous relation, we find
\begin{equation}\label{eq:lambda_to_t2}
  \lambda(\tau) = \frac{1}{32}\Big[ \sqrt{t_2(\tau)(t_2(\tau)+64)} -t_2(\tau) \Big] -2\,.
\end{equation}
We see that $\lambda(\tau)$ is an \textit{algebraic} function of the Hauptmodul
$t_2$. 

Next, let us identify a seed modular form $B_0(\tau)$. As can be checked for
example with SAGE, there is a unique modular form of weight $2$ for $\Gamma_0(2)$ (up to rescaling).
Since $\Gamma(2)\subseteq\Gamma_0(2)$, this form has to be in the space
$\cM_2(\Gamma(2))$, so we can -- using the results from the previous subsection
-- write the ansatz
\begin{equation}
  B_0(\tau)=\textrm{K}(\lambda(\tau))^2 (c_0+c_1\lambda(\tau))\,.
\end{equation}
The coefficients can be fixed by matching $q$-expansions with the expression
delivered by SAGE and one finds that $\cM_2(\Gamma_0(2))$ is generated by
\begin{equation}
  \label{eqn:seedG02}
  B_0(\tau) = \textrm{K}(\lambda(\tau))^2 (\lambda(\tau)-2) \,.
\end{equation}
Equipped with the seed modular form $B_0$, we can now repeat the
analysis from the previous subsection. For a modular form $f(\tau)$ of weight
$2k$ for $\Gamma_0(2)$, the function
\begin{equation}\label{eq:Gamma0(2)_R_def}
  R(t_2(\tau)) = \frac{f(\tau)}{B_0(\tau)^k} 
\end{equation}
is meromorphic and has weight $0$, thus it must be a rational function of the
Hauptmodul $t_2$.
In order to fix the precise form of $R(t_2)$, let us again consider the pole
structure of the right-hand side of eq.~\eqref{eq:Gamma0(2)_R_def}: since both $f(\tau)$ and $B_0(\tau)$ are holomorphic, poles in
$R(\tau)$ can appear only for $B_0(\tau)=0$, which translates into 
\begin{equation}
  \lambda(\tau)=2\quad\text{ or }\quad\textrm{K}(\lambda(\tau))=0\,.
\end{equation}
As spelt out in the previous subsection, the second situation is realised for
$\lambda\to \infty$, i.e., for $\tau\to 1$. Considering this limit, we find
\begin{equation}
  \lim_{\tau\to 1}B_0(\tau)=\lim_{\tau\to 1}\textrm{K}(\lambda(\tau))^2(\lambda(\tau)-2)\sim\lambda(\tau)\Big(\frac{1}{\sqrt{\lambda(\tau)}}\Big)^2=\mathcal{O}(1)\,,
\end{equation}
and we see that $B_0(\tau)$ does not vanish in the limit $\textrm{K}(\lambda(\tau))\to
0$.
As $\textrm{K}(\lambda(\tau))$ is finite for $\lambda(\tau)=2$, $B_0$ will have a simple
zero there. As a function of the Hauptmodul $t_2$, however, $B_0(t_2)$ behaves like
\begin{equation}
  B_0(t_2)\stackrel{t_2\to -64}{\sim}\sqrt{t_2+64}\,,
\end{equation}
which can be seen by expanding eq.~\eqref{eq:lambda_to_t2} around $t_2=-64$. Accordingly,
$R(t_2)$ can at most have a pole of order $\lfloor k/2 \rfloor$ at $t_2=-64$.
Hence, we can write down the following ansatz for $R(t_2)$,
\begin{equation}
  R(t_2)=\frac{P(t_2)}{(t_2+64)^{\lfloor k/2\rfloor}}\,,
\end{equation}
where $P(t_2)$ is a polynomial in the Hauptmodul. Its degree can be
bounded by demanding regularity for $t_2\to \infty$. We obtain in this way the most general form for a modular form of weight $2k$ for
$\Gamma_0(2)$:
\begin{equation}
  \label{eqn:fGam02}
  f(\tau)=\textrm{K}(\lambda(\tau))^{2k}\frac{(\lambda(\tau)-2)^k}{(t_2(\tau)+64)^{\lfloor k/2 \rfloor}}\sum\limits_{m=0}^{\lfloor k/2 \rfloor} c_m t_2(\tau)^m \,.
\end{equation}
In particular we see that 
\begin{equation}
  \dim \cM_{2k}(\Gamma_0(2))=\lfloor k/2 \rfloor +1\,,
\end{equation}
and an explicit basis for $\cM_{2k}(\Gamma_0(2))$ is
\begin{equation}\label{eq:basis_G02}
\textrm{K}(\lambda(\tau))^{2k}\frac{(\lambda(\tau)-2)^k\,t_2(\tau)^m}{(t_2(\tau)+64)^{\lfloor k/2 \rfloor}}\,,\qquad 0\le m\le \lfloor k/2 \rfloor\,.
\end{equation}
We have checked up to weight 10 that our results are in agreement with the explicit basis for modular forms for $\Gamma_0(2)$ obtained by SAGE.
Finally, let us comment on the cusp forms for $\Gamma_0(2)$. $\Gamma_0(2)$ has two cusps, which can be represented by $\tau=i\infty$ and $\tau=0$. The Hauptmodul $t_2$ maps the cusps to
\begin{equation}
t_2(i\infty) = 0 \textrm{~~and~~} t_2(0) = \infty\,.
\end{equation}
We then see from eq.~\eqref{eq:cusp_general} that a basis for $\cS_{2k}(\Gamma_0(2))$ is
\begin{equation}
K(\lambda(\tau))^{2k}\frac{(\lambda(\tau)-2)^k\,t_2(\tau)^m}{(t_2(\tau)+64)^{\lfloor k/2 \rfloor}}\,,\qquad 1\le m\le \lfloor k/2 \rfloor-1\,.
\end{equation}

\begin{exemp}
Since $\Gamma(2)\subseteq \Gamma_0(2)$, we have $\cM_{2k}(\Gamma_0(2)) \subseteq \cM_{2k}(\Gamma(2))$. In particular, this means that we must be able to write every basis element for $\cM_{2k}(\Gamma_0(2))$ in eq.~\eqref{eq:basis_G02} in terms of the basis for $\cM_{2k}(\Gamma(2))$ in eq.~\eqref{eq:basis_Gamma(2)}. Indeed, inserting eq.~\eqref{eq:t2_to_lambda} into eq.~\eqref{eq:basis_G02}, we find,
\begin{equation}\begin{split}
\frac{(\lambda-2)^k\,t_2^m}{(t_2+64)^{\lfloor k/2 \rfloor}}=16^{m-\lfloor k/2 \rfloor}\,\lambda^{2m}\,(1-\lambda)^{\lfloor k/2 \rfloor-m}\,(\lambda-2)^{k-2\lfloor k/2 \rfloor}\,.
\end{split}\end{equation}
It is easy to see that the previous expression is polynomial in $\lambda$ provided that $0\le m\le \lfloor k/2 \rfloor$. Hence, we see that every element in eq.~\eqref{eq:basis_G02} can be written in terms of the basis in eq.~\eqref{eq:basis_Gamma(2)}.
\end{exemp}

\subsection{A basis for modular forms for $\Gamma_0(4)$ and $\Gamma_0(6)$}
In this section we discuss the congruence subgroups $\Gamma_0(4)$ and
$\Gamma_0(6)$. The analysis is identical to the case of $\Gamma_0(2)$ in the
previous section, so we will be brief. There are no modular forms of odd weight
and both groups have genus zero. The respective Hauptmodule $t_4$ and $t_6$ can
be found in ref.~\cite{Maier} in terms of $\eta$-quotients, though their
explicit forms are irrelevant for what follows.  Here we only mention that we can
write the Hauptmodul $t_2$ as a rational function in either $t_4$ or $t_6$~\cite{Maier}
\begin{equation}\label{eq:t2_to_t4_and_t6}
   t_2=t_4(t_4+16) =  \frac{t_6(t_6+8)^3}{t_6+9}\,.
\end{equation}
Since $\Gamma_0(2N)\subseteq \Gamma_0(2)$, the modular form $B_0(\tau)$ in
eq.~\eqref{eqn:seedG02} is a modular form of weight two for $\Gamma_0(2N)$ for
any value of $N$. Hence, we can choose $B_0(\tau)$ as our seed modular form, and so if
$f\in \cM_{2k}(\Gamma_0(2N))$, then $f(\tau)/B_0^k(\tau)$ is is a modular
function for $\Gamma_0(2N)$. In the cases $N=2,3$ which we are interested in
this implies that $f(\tau)/B_0^k(\tau)$ is a rational function in the
Hauptmodul $t_{2N}$,
\begin{equation}
R(t_{2N}(\tau)) = \frac{f(\tau)}{B_0(\tau)^k}\,,\qquad N=4,6\,.
\end{equation}

Let us now analyse the pole structure of $R(t_4)$.  From the last section we
know that $B_0(\tau)$ has a simple zero at $\lambda(\tau)=2$, or equivalently
$t_2=-64$, and eq.~\eqref{eq:t2_to_t4_and_t6} then implies $t_4=-8$. Writing
down an ansatz for $R(t_4)$ and bounding the degree of the polynomial in the
numerator in the usual way, one finds that a basis of modular forms of weight
$2k$ for $\Gamma_0(4)$ is
\begin{equation}
  \label{eqn:fGam04}
\textrm{K}(\lambda(\tau))^{2k}\left(\frac{\lambda(\tau)-2}{t_4(\tau)+8}\right)^k\,t_4(\tau)^m\,,\qquad 0\le m\le k\,.
\end{equation}

$\Gamma_0(4)$ has three cusps which can be represented by
$\tau\in\{i\infty,1,1/2\}$ and which under $t_4$ are mapped to
\begin{equation}
t_4(i\infty)=0\,\qquad t_4(1) = \infty\,,\qquad t_4(1/2) = -16\,.
\end{equation}
Hence a basis for $\cS_{2k}(\Gamma_0(4))$ is
\begin{equation}
\textrm{K}(\lambda(\tau))^{2k}\left(\frac{\lambda(\tau)-2}{t_4(\tau)+8}\right)^k\,t_4(\tau)^m\,(t_4(\tau)+16)\,,\qquad 1\le m\le k-2\,.
\end{equation}

As a last example, let us have a short peek at $\Gamma_0(6)$.
Equation~\eqref{eq:t2_to_t4_and_t6} implies that $B_0(\tau)$ has simple poles for 
\begin{equation}
  t_6(\tau)=-6 \pm 2 \sqrt{3}\,.
\end{equation}
The argument proceeds in the familiar way, with the only difference that now
there are two distinct poles. The most general ansatz for a modular
form of weight $2k$ for $\Gamma_0(6)$ reads
\begin{equation}
  \frac{f(\tau)}{B_0^k(\tau)}=\frac{P(t_6(\tau))}{[(t_6(\tau)+6-2\sqrt{3})(t_6(\tau)+6+2\sqrt{3})]^k}
  =\frac{P(t_6(\tau))}{(t_6(\tau)^2+12t_6(\tau)+24)^k}\,,
\end{equation}
where the degree of the polynomial $P$ can again be bounded by the common
holomorphicity argument. This leads to the following basis for modular
forms of weight $2k$ for $\Gamma_0(6)$, 
\begin{equation}
  \label{eqn:fGam06}\textrm{K}(\lambda(\tau))^{2k}\left(\frac{\lambda(\tau)-2}{t_6(\tau)^2+12t_6(\tau)+24}\right)^k\,t_6(\tau)^m\,,\qquad 0\le m\le 2k\,.
\end{equation}
The cusps of $\Gamma_0(6)$ are represented by $\tau\in\{i\infty,1,1/2,1/3\}$,
or equivalently
\begin{equation}
t_6(i\infty)=0\,,\qquad t_6(1)=\infty\,,\qquad t_6(1/2) = -8\,,\qquad t_6(1/3) = -9\,.
\end{equation}
Hence a basis for $\cS_{2k}(\Gamma_0(6))$ is, with $1\le m\le 2k-3$,
\begin{equation}
\textrm{K}(\lambda(\tau))^{2k}\left(\frac{\lambda(\tau)-2}{t_6(\tau)^2+12t_6(\tau)+24}\right)^k\,t_6(\tau)^m\,(t_6(\tau)+8)\,(t_6(\tau)+9)\,.
\end{equation}


\subsection{A basis for modular forms for $\Gamma_1(6)$}
\label{ssec:modformG16}

As a last application we discuss the structure of modular forms for
$\Gamma_1(6)$, which is known to be relevant for the sunrise and kite
integrals~\cite{Bloch:2013tra,Adams:2017ejb}. The general story will be very
similar to the examples in previous sections. In particular, $\Gamma_1(6)$ has
genus zero, and $\Gamma_1(6)$ and  $\Gamma_0(6)$ have the same Hauptmodul
$t_6$~\cite{Bloch:2013tra}. Here we find it convenient to work with an
alternative Hauptmodul $t$ which is related to $t_6$ by a simple M\"obius
transformation~\cite{Adams:2017ejb},
\begin{equation}
  \label{eqn:tt6}
t = \frac{t_6}{t_6+8}\,.
\end{equation}
The main difference to the previous examples lies in the fact that
$\left(\begin{smallmatrix}-1&0\\0&-1\end{smallmatrix}\right)\notin\Gamma_1(6)$,
and so $\Gamma_1(6)$ admits modular forms of odd weight. In particular, it is
known that $\cM_1(\Gamma_1(6))$ is two-dimensional (this can easily be checked
with SAGE for example). Therefore, we would like to choose our seed modular
form to have weight one. We find it convenient to choose as seed modular form a
solution of the Picard-Fuchs operator associated to the sunrise
graph~\cite{Laporta:2004rb,Bloch:2013tra}. A particularly
convenient choice is
\begin{equation}
B_1(\tau) = \Psi_1(t(\tau))\,,
\end{equation}
where
\begin{equation}\label{eq:psi1_def}
\Psi_1(t) = \frac{4}{[(t-9)(t-1)^3]^{1/4}}\,\textrm{K}\left(\frac{t^2-6t-3+\sqrt{(t-9)(t-1)^3}}{2\sqrt{(t-9)(t-1)^3}}\right)\,.
\end{equation}
It can be shown that $\Psi_1(t(\tau))$ is indeed a modular form of weight one
for $\Gamma_1(6)$~\cite{Adams:2017ejb}.

Next consider a modular form $f(\tau)$ of weight $k$ for $\Gamma_1(6)$.
Following the usual argument, the ratio
\begin{equation}
R(t(\tau)) = \frac{f(\tau)}{B_1(\tau)^k}
\end{equation}
is a rational function in the Hauptmodul $t$ with poles at most at points where
$\Psi_1(t)$ vanishes. It is easy to check that the only zero of $\Psi_1(t)$ is
at $t=\infty$, and we have
\begin{equation}
  \Psi_1(t)\stackrel{t\to \infty}{\sim}1/t\,.
\end{equation}
Hence, $R(t)$ must be a polynomial in $t$ whose degree is bounded by requiring
that $\Psi_1(t)^k\,R(t)$ be free of poles at $t=\infty$. It immediately follows
that a basis of modular forms of weight $k$ for $\Gamma_1(6)$ is
\begin{equation}\label{eqn:fGam16}
\Psi_1(t(\tau))^k\,t(\tau)^m\,,\qquad 0\le m\le k\,.
\end{equation}
The cusps of $\Gamma_1(6)$ can be represented by
$\tau\in\{i\infty,1,1/2,1/3\}$, and they are mapped to
\begin{equation}
t(i\infty)=0,,\qquad t(1)=1\,,\qquad t(1/2) = \infty\,,\qquad t(1/3) = 9\,.
\end{equation}
So a basis of cusp forms of weight $k$ for $\Gamma_1(6)$ is
\begin{equation}
\Psi_1(t(\tau))^k\,t(\tau)^m\,(t(\tau)-1)\,(t(\tau)-9)\,,\qquad 1\le m\le k-3\,.
\end{equation}

Let us conclude by commenting on the structure of the modular forms for $\Gamma_1(6)$, and their relationship to modular forms for $\Gamma_0(6)$. Since $\Gamma_1(6)\subseteq \Gamma_0(6)$ we obviously have $\cM_k(\Gamma_0(6))\subseteq\cM_k(\Gamma_1(6))$. Moreover, from eq.~\eqref{eqn:fGam06} and~\eqref{eqn:fGam16} we see that for even weights these spaces have the same dimension, and so we conclude that
\begin{equation}
\cM_{2k}(\Gamma_1(6)) = \cM_{2k}(\Gamma_0(6))\,.
\end{equation}
There is a similar interpretation of the modular forms of odd weights. It can be shown that the algebra of modular forms for $\Gamma_1(N)$ admits the decomposition
\begin{equation}\label{eq:G1N_decomposition}
\cM_k(\Gamma_1(N)) = \bigoplus_{\chi}\cM_k(\Gamma_0(N),\chi)\,,
\end{equation}
where the sum runs over all Dirichlet characters modulo $N$, i.e., all homomorphisms $\chi :\mathbb{Z}_N^\times\to \bc^\times$. Here $\cM_k(\Gamma_0(N),\chi)$ denotes the vector space of modular forms of weight $k$ for $\Gamma_0(N)$ with character $\chi$, i.e., the vector space of holomorphic functions $f:\overline{\bh}\to\bc$ such that
\begin{equation}
f\left(\frac{a\tau+b}{c\tau+d}\right) = \chi(d)\,(c\tau+d)^k\,f(\tau)\,,\qquad \abcd\in\Gamma_0(N)\,.
\end{equation}
For $N=6$ there are two Dirichlet characters modulo 6, 
\begin{equation}
\chi_0(n) =1 \qquad \textrm{~~and~~}\qquad \chi_1(n) = (-1)^n\,.
\end{equation}
Hence, in the case we are interested in, eq.~\eqref{eq:G1N_decomposition} reduces to
\begin{equation}
\cM_k(\Gamma_1(6))  = \cM_k(\Gamma_0(6),\chi_0)  \oplus \cM_k(\Gamma_0(6),\chi_1) = \cM_k(\Gamma_0(6)) \oplus \cM_k(\Gamma_0(6),\chi_1)\,.
\end{equation} 
We then conclude that
\begin{equation}
 \cM_{2k}(\Gamma_0(6),\chi_1) = 0\quad \textrm{~~and~~} \quad \cM_{2k+1}(\Gamma_0(6),\chi_1) =  \cM_{2k+1}(\Gamma_1(6))\,.
 \end{equation}

\vspace{1mm}\noindent



\section{Some examples and applications}
\label{sec:application}
%

\subsection{Elliptic multiple zeta values as iterated integrals over modular
forms for $\Gamma(2)$}

Elliptic multiple zeta values have appeared in calculations in quantum
field theory and string theory in various formulations during the last couple
of years. While initially formulated as special values of elliptic multiple
polylogarithms, they can be conveniently rewritten as iterated integrals over
the Eisenstein series $G_{2k}$ defined in eq.~\eqref{eqn:eis}
\cite{Broedel:2015hia}.  In other words, elliptic multiple zeta values are
iterated integrals over modular forms for $\Gamma(1)=\slz$ (though it is known
that not every such integral defines an element in the space of elliptic
multiple zeta value~\cite{BrownMMV}).

We have seen in Example~\ref{ex:seven} that every modular form for $\Gamma(1)$
is a modular form for $\Gamma(2)$. In particular, for $k>1$ we can always write
$G_{2k}$ as the $2k$-th power of $\textrm{K}(\lambda(\tau))$ multiplied by a polynomial
$\cG_{2k}$ of degree $k$ in $\lambda(\tau)$ (see eq.~\eqref{eq:Eis_to_G}). The
case $k=1$ is special, and involves the elliptic integral of the second kind,
see eq.~\eqref{eq:G2_def}.

As a consequence, we can write every iterated integral of Eisenstein series of
level $N=1$, and thus every elliptic multiple zeta value, as iterated integrals
over integration kernels that involve powers of complete elliptic
integrals of the first kind multiplied by the polynomials $\cG_{2k}(\lambda(\tau))$.
More precisely, consider the one-forms $d\tau\,G_{2k}(\tau)$ which define
iterated integrals of Eisenstein series of level one. Changing variables from
$\tau$ to $\ell=\lambda(\tau)$, we obtain, for $k>1$,
\begin{equation}\label{eq:eMZV_kernel}
d\tau\,G_{2k}(\tau) =  \frac{i \pi \,d\ell}{4 \,\ell\,(\ell-1)}\,\textrm{K}(\ell)^{2k-2}\,\cG_{2k}(\ell)\,,
\end{equation}
where Jacobian is given by
\begin{equation}\label{eq:jacobian}
  \quad 2 \pi \, i \partial_\tau \lambda(\tau) = 8 \lambda(\tau) (\lambda(\tau)-1) \textrm{K}(\lambda(\tau))^2\,. 
\end{equation}
Note that we also need to include the Eisenstein series of weight zero,
$G_0(\tau)=-1$, and eq.~\eqref{eq:eMZV_kernel} remains valid if we let
$\cG_0(\ell)=-1$. For $k=1$ we can derive from eq.~\eqref{eq:G2_def} a similar
relation involving the complete elliptic integral of the second kind.  As a
conclusion, we can always write iterated integrals of Eisenstein series of
level one in terms of iterated integrals involving powers of complete elliptic
integrals multiplied by rational functions. We stress that this construction is
not specific to level $N=1$ or to Eisenstein series, but using the results from
previous sections it is possible to derive similar representations of
`algebraic type' for iterated integrals of general modular forms.

\subsection{A canonical differential equation for some classes of hypergeometric functions}
\label{sec:2F1}

As an example of how the ideas from previous sections can be used in the context of differential equations, let us consider
the family of integrals
\begin{align}
T(n_1,n_2,n_3) = \int_0^1 dx\, x^{-1/2 + n_1 + a\, \epsilon} (1-x)^{-1/2 + n_2 + b\, \epsilon} (1-z\, x)^{-1/2 + n_3 + c\,\epsilon}\,.
\label{eq:2F1}
\end{align}
This family is related to a special class of hypergeometric functions whose
$\epsilon$-ex\-pansion has been studied in detail in refs.~\cite{Broedel:2017kkb,Broedel:2018iwv}.
It is easy to show that all integrals in eq.~\eqref{eq:2F1}, for any choice
of $n_1,n_2,n_3$, can be expressed as linear combination of two independent
master integrals, which can be chosen as
\begin{equation}
F_1 = T(0,0,0) \textrm{~~~and~~~} \quad F_2 = T(1,0,0)\,.
\end{equation} 
The two masters satisfy the system of two differential equations,
\begin{align}
\partial_z\, F = (A + \epsilon B) F\,, \quad \mbox{with} \quad F = \left(  F_1 , F_2 \right)^T \,,\label{eq:deqs2F1}
\end{align}
where $A,B$ are two $2 \times 2$ matrices
\begin{align}
A &= \frac{1}{z} \left( \begin{array}{ccc} 0 && 0 \\ 1/2 && -1 \end{array}  \right) + \frac{1}{z-1} \left( \begin{array}{ccc} -1/2 && 1/2 \\ -1/2 && 1/2  \end{array} \right)\,, \\
B &= \frac{1}{z} \left( \begin{array}{ccc} 0 && 0 \\ a && -a-b  \end{array} \right) + \frac{1}{z-1} \left( \begin{array}{ccc} -a && a+b+c \\ -a && a+b+c  \end{array} \right)\,.
\end{align}
A suitable boundary condition for the differential equations~\eqref{eq:deqs2F1}
can be determined by computing directly the integrals in eq.~\eqref{eq:2F1} at
$z=0$
\begin{align}
\lim_{z \to 0}F = \frac{\Gamma \left(a \epsilon +\frac{1}{2}\right) \Gamma \left(b \epsilon +\frac{1}{2}\right)}{\Gamma ( 1 + (a +b )\epsilon )}
\left( 1, \frac{2 a \epsilon +1}{2 \epsilon  (a+b)+2} \right)^T\,. \label{eq:bound}
\end{align}

We are now ready to solve the differential equations.  It is relatively easy to
see that by performing the following change of basis
\begin{align}
F = M G\,, \quad G = (G_1, G_2)^T\,,
\end{align}
with
%
\begin{align}
M = \frac{1}{(2 (a+b+c) \epsilon +1)}\left(
\begin{array}{ccc}
 2 K(z) (2 (a+b+c) \epsilon +1) & & 0 \\ \\
 \frac{\epsilon }{2 z K(z)}
 -\frac{2 E(z)}{z }
 +\frac{2 ((a+b) \epsilon +(a+c) z \epsilon +1) K(z)}{z } & &
   \frac{\epsilon }{2 \,z \, K(z)} 
\end{array}
\right)\,,
\end{align}
the new master integrals $G_1,G_2$ fulfil the system of differential equations
\begin{align}
\partial_z G = \frac{\epsilon}{2\,z \, (z-1)\, K(z)^2}\, \Omega\, G \,,\label{eq:neweq}
\end{align}
where the matrix $\Omega$ can be written as
\begin{align}
\Omega = \Omega_0 + \Omega_1 + \Omega_2\,,
\end{align}
with 
\begin{align}
&\Omega_0 = \frac{1}{4} \left(
\begin{array}{cc}
 1 & 1 \\
 -1 & -1 \\
\end{array}
\right) \,, \qquad 
\Omega_1 =  (a+b+(c-a) z) \, K(z)^2 \left(
\begin{array}{cc}
1 & 0 \\
 0 & 1 \\
\end{array}
\right)\,,\nonumber \\ 
&\Omega_2 = 4 \,  \left((a+b)^2+(a+c)^2 z^2-2 \left(a^2+b a+c a-b c\right) z\right) \, K(z)^4\, \left(
\begin{array}{cc}
 0 & 0 \\
 1 & 0 \\
\end{array}
\right)\,.
\end{align}
We stress that the differential equations in eq.~\eqref{eq:neweq} are
$\epsilon$-factorised.

In order to {solve} eq.~\eqref{eq:neweq}, let us change
variable from $z$ to $\tau$ via $z =
\lambda(\tau) $, where $\lambda$ denotes the modular $\lambda$-function. Using the form of the Jacobian in eq.~\eqref{eq:jacobian}, 
we find that the differential equations become
\begin{align}
\partial_\tau G = \frac{2\, \epsilon}{\pi \,i}\, \Omega\, G\,. \label{eq:neweq2}
\end{align}

As the last step, we know from the discussion in Section~\ref{eq:sec_Gamma(2)} that a
basis of modular forms of weight $2k$ for $\Gamma(2)$ is given by
$\lambda(\tau)^p K(\lambda(\tau))^{2 k}$, with $0\leq p \leq k$.  Using this,
we see that the entries of $\Omega$ are indeed linear combinations of modular
forms of $\Gamma(2)$.
The boundary condition at $z=0$ in eq.~\eqref{eq:bound} translates directly into a boundary condition in $\tau =
i \infty$. Hence, we have proved that the two entries of the vector $G$ can be written, to
all orders in $\epsilon$, in terms of iterated integrals of modular forms for
$\Gamma(2)$.

\subsection{Modular forms for $\Gamma_1(6)$ and the sunrise and the kite integrals}

In section 3 of ref.~\cite{Adams:2018yfj} the integral family for the integral
for the kite family has been investigated, and it was shown that all the kernels
presented in eq.~(34) of ref.~\cite{Adams:2018yfj} are modular forms for the congruence
subgroup $\Gamma_1(6)$. The analysis of ref.~\cite{Adams:2018yfj} relies on a direct matching of the kernels that appear in the sunrise and kite integrals
to the basis of Eisenstein for $\Gamma_1(6)$ given in the mathematics literature. 
In Section~\ref{ssec:modformG16} we have constructed an alternative basis for $\Gamma_1(6)$, and so we must be able to write all the integration kernels that appear in the sunrise integral in terms of our basis. This is the content of this section, and we argue that our basis makes the fact that the sunrise and kite integrals can be expressed in terms of iterated integrals of modular forms for $\Gamma_1(6)$ completely manifest.

In order to make our point, we proceed by example, and we consider in particular the function $f_2$ defined in eq.~(34) of ref.~\cite{Adams:2018yfj}. This function is one of the coefficients that appear in the differential equation satisfied by the master integrals of the kite topology, after the differential equations have been transformed to $\epsilon$-form~\cite{Adams:2018yfj,Henn:2013pwa}. All other coefficients appearing in the system of differential equations can be analysed in the same way. The function $f_2$ is defined as
\begin{equation}
  \label{eqn:Wf2}
  f_2(x)=\frac{1}{24\pi^2}\Psi_1(x)^2\big(3x^2 - 10 x - 9\big)
\end{equation}
where  $x=p^2/m^2$, with $m$ the mass of the massive state flowing in the loop and $p$ the external momentum, and (in our notations) $\Psi_1$ was defined in eq.~\eqref{eq:psi1_def} (note that compared to ref.~\cite{Adams:2018yfj} we have explicitly inserted the expression for the Wronskian $W$ as a function of $x$ into the definition of $f_2$). 
From the form of eq.~\eqref{eqn:Wf2} we can immediately read off that $f_2$ defines a modular form for $\Gamma_1(6)$. Indeed, changing variables to $x=t(\tau)$, where $t(\tau)$ is the Hauptmodul for $\Gamma_1(6)$ introduced in Section~\ref{ssec:modformG16}, we see that $f_2(t(\tau))$ takes the form
$\Psi_1(t(\tau))^2\,P(t(\tau))$, where $P$ is a polynomial of degree two. Thus $f_2(t(\tau))$ can be written as a linear combination of the basis of modular forms of weight two for $\Gamma_1(6)$ given in eq.~\eqref{eqn:fGam16}, and so $f_2(t(\tau))$ itself defines a modular form of weight two for $\Gamma_1(6)$. It is easy to repeat the same analysis for all the coefficients that appear in the system of differential equations for sunrise and kite integrals, and we can conclude that the sunrise and kite integrals can be written in terms of iterated integrals of modular forms to all orders in $\epsilon$. We emphasise that we have reached this conclusion solely based on the knowledge of the Hauptmodul of $\Gamma_1(6)$ and the fact that $\Psi_1(t(\tau))$ defines a modular form of weight one for $\Gamma_1(6)$. The rest follows from our analysis performed in Section~\ref{ssec:modformG16}, and we do not require any further input from the mathematics literature on the structure of modular forms for $\Gamma_1(6)$.
%
%
%

\vspace{1mm}\noindent



\section{Conclusions and Outlook}
\label{sec:X}

In this contribution to the proceedings of the conference ``Elliptic
integrals, elliptic functions and modular forms in quantum field theory'', we presented a systematic way of writing a basis modular forms for congruence subgroups of the modular group $\slz$ in terms of powers of complete elliptic integrals of the first kind multiplied by algebraic functions. We considered congruence groups whose modular curves have genus zero and as such all modular forms can be written as powers of complete elliptic integrals of the first kind multiplied by rational functions of their corresponding Hauptmodule.
Our construction relied simply on the knowledge of a seed modular form of lowest weight for each congruence group and its analytic properties. This, put together with the holomorphicity condition for modular forms, allowed us to write a general ansatz for a basis of modular forms.

We presented concrete examples for the congruence groups $\Gamma(2),\,\Gamma_0(N)$ for $N=2,4,6$, and finally $\Gamma_1(6)$ which features in physical applications such as the sunrise and kite integrals. By this method we showed how to write elliptic multiple zeta values as iterated integrals of rational functions weighted by complete elliptic integrals. Likewise, rewriting the differential equations of the sunrise and kite integrals, we were able to show that to all orders in $\varepsilon$ these can be written as iterated integrals of modular forms for $\Gamma_1(6)$, confirming the findings of \cite{Adams:2017ejb,Adams:2018yfj}.

We hope that our construction constitutes a first step into clarifying
the connection between solutions of differential equations for elliptic Feynman integrals and elliptic multiple polylogarithms,
allowing for a systematic application of this class of functions to 
realistic physical problems.


\vspace{1mm}\noindent

\vspace*{4mm}
\noindent
{\bf Acknowledgment.} We would like to thank the ``Kolleg Mathematik und Physik
Berlin'' for supporting the workshop ``Elliptic  integrals, elliptic functions
and modular forms in quantum field theory''. 
This research was supported by the the ERC grant 637019 ``MathAm'', and the U.S.
Department of Energy (DOE) under contract DE-AC02-76SF00515.
\bibliography{bib}
\end{document}